\documentclass[aps,pra,twocolumn,showpacs,superscriptaddress,floatfix,10pt]{revtex4-2}
\usepackage{framed}
\usepackage{graphicx}
\usepackage{xfrac}
\usepackage{amsmath}
\usepackage{epsfig}
\usepackage{helvet}
\usepackage{amssymb}
\usepackage{soul}
\usepackage{xcolor}
\usepackage{framed}
\usepackage{hyperref}
\usepackage{amsthm}

\newcommand{\CC}{C\nolinebreak\hspace{-.05em}\raisebox{.4ex}{\tiny\bf +}\nolinebreak\hspace{-.10em}\raisebox{.4ex}{\tiny\bf +}}

\newtheoremstyle{mine}
  {\topsep}
  {\topsep}
  {\itshape}
  {0pt}
  {\bfseries}
  {:}
  {5pt plus 1pt minus 1pt}
  {}
\theoremstyle{mine}

\begin{document}

\title{Entropy Inequalities Constrain Holographic Erasure Correction}

\author{Bart{\l}omiej Czech and Sirui Shuai and Yixu Wang}
\affiliation{Institute for Advanced Study, Tsinghua University, Beijing 100084, China}
\vskip 0.25cm

\begin{abstract}
\noindent
We interpret holographic entropy inequalities in terms of erasure correction. The non-saturation of an inequality is a necessary condition for certain schemes of holographic erasure correction, manifested in the bulk as non-empty overlaps of corresponding entanglement wedges.
\end{abstract}

\maketitle
\textit{Introduction.---} 
The AdS/CFT correspondence \cite{adscft}---probably the best understood UV-complete theory of gravity---is a duality between gravity with asymptotically anti-de Sitter boundary conditions (AdS) and a conformal field theory (CFT) living on its asymptotic boundary. One key element of the duality is the Ryu-Takayanagi proposal \cite{rt1, rt2, hrt}, which under the simplest circumstances posits that the entanglement entropy of a boundary (CFT) subsystem $A$ equals the minimal area of a bulk (AdS) surface homologous to $A$. Another key fact about the duality is that the physics in the bulk region between $A$ and its homologous minimal surface, called the entanglement wedge $E(A)$, can be recovered from $A$ alone---a statement known as subregion duality \cite{subregion1, subregion2}. These assertions impose a degree of locality on the holographic (bulk-to-boundary) map.

The holographic map has interesting properties with respect to set inclusion. Select a number of {disjoint} subsystems $A_s$ on the boundary. The entanglement wedge nesting theorem (EWN) \cite{maximin} says that the wedges of two non-overlapping subsystems are disjoint:
\begin{equation}
A_s \cap A_t = \emptyset \quad \Rightarrow \quad E(A_s) \cap E(A_t) = \emptyset
\label{eqewn}
\end{equation}
However, a similar statement where three subsystems have no common overlap does not hold, for example \cite{errorcorr}
\begin{equation}
E(A_1 A_2) \cap E(A_2 A_3) \cap E(A_3 A_1) \neq \emptyset~~\textrm{is possible} \label{exampleerror}
\end{equation}
even though $A_1 A_2 \cap A_2 A_3 \cap A_3 A_1 = \emptyset$. (Here and in the following $A_1 A_2 \equiv A_1 \cup A_2$.) Reference~\cite{errorcorr} traced the origin of phenomenon~(\ref{exampleerror}) to erasure-correcting properties of the holographic map. In this view, bulk physics comprises logical degrees of freedom, which are redundantly encoded in the CFT. If a limited portion of the CFT data is erased (e.g. erasing $A_3$ to retain $A_1 A_2$), the retained data may suffice to recover the bulk information. The erasure-correcting aspects of the bulk-to-boundary map have been extensively studied \cite{rtcorr}, including explicit tensor network models \cite{happy, rtn}.

The Ryu-Takayanagi proposal also has a noteworthy consequence: it imposes conditions on the entanglement entropies of the boundary quantum state. Suppose the boundary is divided into $2k + 1$ subregions $A_s$, with the indices interpreted modulo $2k+1$ ($A_{2k+2} \equiv A_1$ etc.) If the quantum state is holographically dual to a semiclassical geometry and the Ryu-Takayanagi proposal holds then the entanglement entropies satisfy \cite{mmiref, hec}:
\begin{equation}
\!\!
\sum_{s=1}^{2k+1} \big( S_{A_s A_{s+1} \ldots A_{s+k}} - S_{A_{s} \ldots A_{s+k-1}} \big) \geq S_{A_1 A_2 \ldots A_{2k+1}}
\,\,\,\label{cyclic}
\end{equation}
These conditions eliminate certain entanglement structures, such as the GHZ-type entanglement on four or more constituents \cite{ghz}, from representing semiclassical bulk geometries. Beyond (\ref{cyclic}), we currently know two infinite families of holographic entropy inequalities \cite{yunfei, ourineqs, facetness} (see~(\ref{eqtoric}) below and Supplemental Material \cite{sm}) and 1868 isolated inequalities \cite{cuenca, 1866}, which do not seem to conform to any discernible pattern.

The multitude of the inequalities, and the lack of a known principle to organize them, suggest a question: Can we interpret the holographic entropy inequalities in physical, ideally operational terms? The inequalities express inviolable properties of Ryu-Takayanagi surfaces, which in turn (by subregion duality) determine which bulk regions are protected against which boundary erasures. It is therefore natural to suspect that a physical interpretation of the inequalities should concern erasure correction. 
This paper offers such an interpretation.
\smallskip

\textit{Example.---} Take inequality~(\ref{cyclic}) with $k=2$. Imagine designing a holographic erasure correcting code where five parties $A_s$ ($1 \leq s \leq 5$; $s \equiv s+5$), possibly in the presence of a sixth party $O$, encode some logical information. The goal is to ensure that the logical information be recoverable from $A_1 A_2 A_3$ and from $A_3 A_4 A_5$ but \emph{not} from $A_2 A_3 A_4$ nor $A_4 A_5 A_1$ nor $A_5 A_1 A_2$. As an illustration of our more general claim, we posit that if inequality~(\ref{cyclic}) is saturated then this goal cannot be achieved.

\begin{figure}
    \centering
$\begin{array}{cp{0.08\linewidth}c}
    \includegraphics[width=0.437\linewidth]{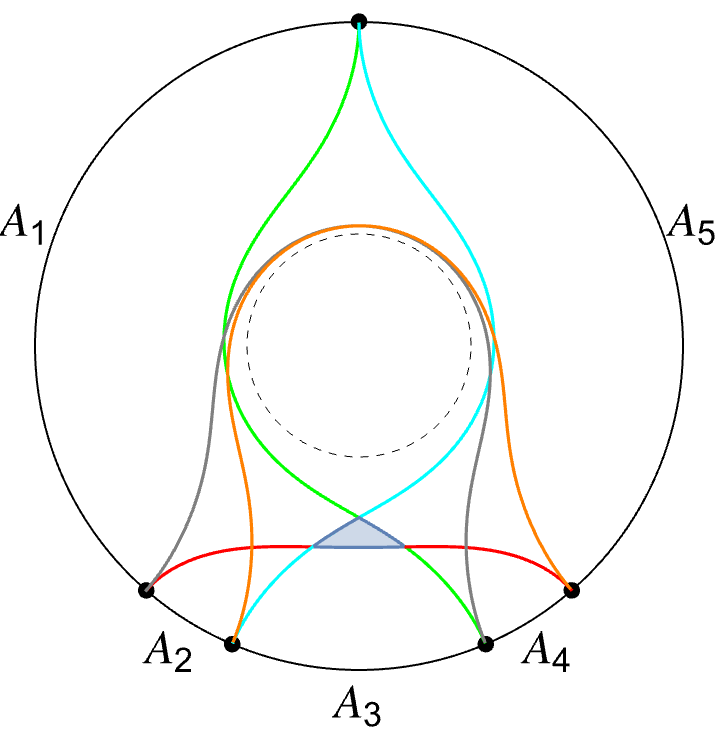} & &
    \includegraphics[width=0.413\linewidth]{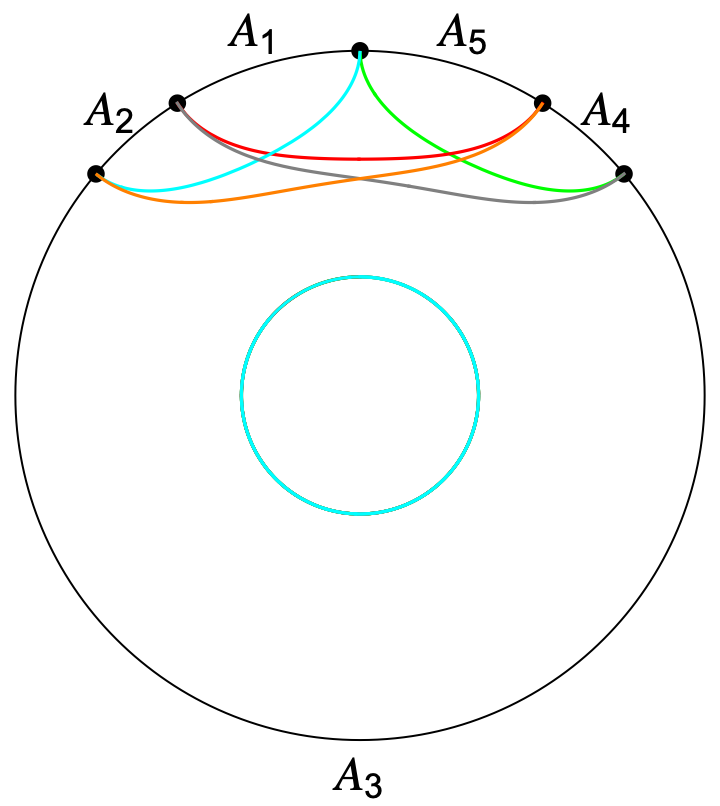}\end{array} $
    \includegraphics[width=0.99\linewidth]{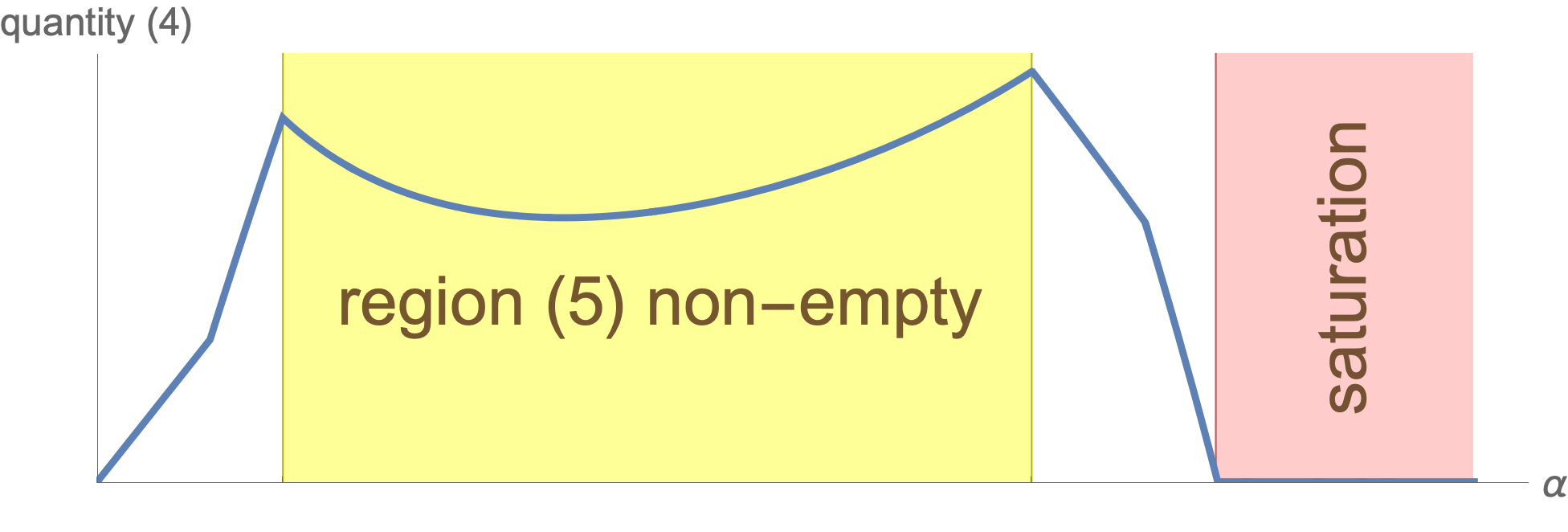}
    \caption{Example~(\ref{5partyex}-\ref{region5}) applied to the static BTZ black hole: we vary $A_3 = (-\alpha, \alpha)$ and keep $A_2, A_4$ of fixed small size. Quantity~(\ref{5partyex}) is non-zero and the highlighted region~(\ref{region5}) is non-empty only if the Ryu-Takayanagi surfaces are in the phases shown. As claimed, the two circumstances never coincide.}
    \label{fig:51example}
\end{figure}

Because bulk geometries encapsulate recoverable logical information in the form of entanglement wedges, our claim is really a statement about entanglement wedges; see Figure~\ref{fig:51example}. We can transcribe it explicitly as follows:
\begin{align}
\!\!\!\sum_{1\leq s \leq 5} S_{A_s^+} & - \sum_{1\leq s \leq 5} S_{A_s^-} - S_{A_1 A_2 A_3 A_4 A_5} = 0
\quad \Rightarrow \label{5partyex} \\
E(A_1^+) & \cap \overline{E(A_2^+)} \cap E(A_3^+) \cap \overline{E(A_4^+)} \cap \overline{E(A_5^+)} = \emptyset
\label{region5}
\end{align}
Here we introduce the $k$-dependent definitions
\begin{align}
A_s^+ &= A_s A_{s+1} \ldots A_{s+k-1} A_{s+k} \label{defap} \\
A_s^- &= A_s A_{s+1} \ldots A_{s+k-1} \label{defam}
\end{align}
for consecutively labeled combinations of subsystems $A_s$, which appear in inequalities~(\ref{cyclic}). By comparing with statement~(\ref{exampleerror}), we see that ours is a \emph{sufficient} condition for when a certain type of erasure correction does \emph{not} work. By contrapositive, the \emph{non}-saturation of a holographic inequality is a \emph{necessary} condition for certain schemes of erasure correction. 

Figure~\ref{fig:51example} presents one instance of~(\ref{5partyex}-\ref{region5}). Five parties $A_s$ control consecutive intervals on the circular boundary of a static BTZ black hole \cite{btz}; the sixth party $O$ represents the purifying system, i.e. the second asymptotic region of the Kruskal-extended spacetime \cite{juaneternalAdS}. We vary $A_3 = (-\alpha, \alpha)$ while keeping the sizes of $A_1, A_5$ and $A_2, A_4$ (the latter fixed and small) pairwise equal. As claimed, the bulk region in question is non-empty only when the inequality is not saturated.
\smallskip

\textit{Preliminaries.---}  To state our claim in full generality and to prove it, we need the concept of a \emph{contraction map} \cite{hec}. Exhibiting a contraction that obeys appropriate boundary conditions is how all the presently known holographic entropy inequalities were proven. It is not known whether every holographic entropy inequality can be proven using a contraction (but see \cite{ning1, ning2}) and whether this method of proof is valid when the bulk geometry and its dual quantum state are time-dependent (but see \cite{withxi, mattveronika, mattveronikanew}). We sidestep both complications by considering only inequalities for which a contraction proof exists, and only working with holographic erasure correcting codes that are time reversal-symmetric. 

Consider a schematic inequality
\begin{equation}
\sum_{i=1}^l \alpha_i\, S_{X_i} \geq \sum_{j=1}^r \beta_j\, S_{Y_j}
\label{schematic}
\end{equation}
where all coefficients $\alpha_i$ and $\beta_j$ are positive. Under the stated assumptions, the minimal surface that computes $S_{X_i}$ divides the time-reversal symmetric bulk slice into two halves: $\tilde{E}(X_i)$ that lives within $E(X_i)$, and its complement. The entanglement wedge $E(X_i)$ is the bulk domain of dependence of $\tilde{E}(X_i)$ \cite{subregion1}, but in what follows we sloppily refer to $\tilde{E}(X_i)$ as `the entanglement wedge of $A$' and drop the tilde from the notation. This avoids clutter and causes no confusion because all our analysis happens on the time-reversal symmetric bulk slice. 

Taking intersections of $E(X_i)$ and $\overline{E(X_i)}$ over all terms on the `greater-than' side (LHS) of the inequality, we partition the bulk slice into $2^l$ regions called $W(x)$. Here $x \in \{0,1\}^l$ is a bit string, which summarizes whether $W(x)$ is inside ($x_i = 1$) or outside ($x_i = 0$) the entanglement wedges $E(X_i)$. By subregion duality, we can view $x$ as a list of which boundary subsystems can ($x_i = 0$) or cannot ($x_i = 1$) be erased while preserving the bulk information in $W(x)$. For instance, the pattern of erasure correction featured in~(\ref{region5}) is $W(10100)$. 

To prove (\ref{schematic}), we combine regions $W(x)$ into \emph{candidate} entanglement wedges of subsystems $Y_j$ such that:
\begin{equation}
\sum_{i=1}^l \alpha_i\, S_{X_i} 
~\stackrel{(\flat)}{\geq}~ \sum_{j=1}^r \beta_j\, S_{Y_j}^{\rm candidate} 
~\stackrel{(\sharp)}{\geq}~ \sum_{j=1}^r \beta_j\, S_{Y_j}
\label{schematic2}
\end{equation}
Because the physically realized entanglement entropies of $Y_j$ are minimal, step $(\sharp)$ in (\ref{schematic2}) is automatic. The clou of the proof is to select the right combinations of $W(x)$ to form candidate entanglement wedges of $Y_j$ such that $S_{Y_j}^{\rm candidate}$ automatically satisfy step $(\flat)$. That selection is encoded in a contraction map $f$. 

The map takes bit strings $x \in \{0,1\}^l$ into bit strings $f(x) \in \{0,1\}^r$. From the definition of $W(x)$, observe that (the time slice of) the entanglement wedge of $X_i$ is the union of all regions $W(x)$ such that the $i^{\rm th}$ bit of $x$ is 1:
\begin{equation}
E(X_i) = \cup_{x | x_i = 1} \, W(x)
\label{exi}
\end{equation}
To form a candidate entanglement wedge of $Y_j$, we write an analogous expression with $f(x)$:
\begin{equation}
E(Y_j)^{\rm candidate} = \cup_{x | f(x)_j = 1} \, W(x)
\label{eyjcandidate}
\end{equation}
To prove~(\ref{schematic}), map $f$ must satisfy two conditions:
\begin{itemize}
\item The contraction property: 
\begin{equation}
\sum_{i=1}^l \alpha_i \, |x_i - \tilde{x}_i| \geq \sum_{j=1}^r \beta_j \, |f(x)_j - f(\tilde{x})_j|
\label{defcontraction}
\end{equation}
for all $x, \tilde{x} \in \{0,1\}^l$. This validates step $(\flat)$ in (\ref{schematic2}).
\item Boundary conditions: For each indivisible subsystem $A$ (in example (\ref{5partyex}) this includes the five subsystems $A_s$ and the purifier $O$), form indicator bit strings $x^A$ and $f^A$. The $i^{\rm th}$ bit of $x^A$ is 1 if $A \subset X_i$ and 0 otherwise, and likewise for $f^A$ with respect to the $Y_j$. For all indivisible subsystems $A$ we require:
\begin{equation}
f(x^A) = f^A
\label{defbc}
\end{equation}
This ensures that the candidate entanglement wedges~(\ref{eyjcandidate}) satisfy the homology condition in the Ryu-Takayanagi proposal. 
\end{itemize}
How these two conditions achieve their goals has been reviewed in many references, e.g.~\cite{hec, withxi, review, qeshec, viols}.
\smallskip

\textit{Candidate entanglement wedges violate entanglement wedge nesting.---} In eq.~(\ref{eqewn}) we highlighted a general feature of entanglement wedges---that disjoint boundary subsystems have disjoint wedges. If we work with pure states---which we can always do by introducing a purifying boundary system $O$---then $\overline{E(X_i)} = E(\overline{X_i})$ and the same fact can be expressed in another way:
\begin{equation}
X \subset Y \quad \Rightarrow \quad E(X) \subset E(Y)
\label{eqewn2}
\end{equation}
This property, called entanglement wedge nesting, is necessary for subregion duality to hold or else tracing out $Y \setminus X$ would enhance our ability to reconstruct the bulk.

While physically realized entanglement wedges always obey~(\ref{eqewn}) and (\ref{eqewn2}), the candidate entanglement wedges employed in proofs by contraction need not do so. In fact, as two of us recently pointed out \cite{viols}, the exact opposite is the case: properties (\ref{defcontraction}) and (\ref{defbc}) of contraction maps directly imply that the candidate wedges defined in (\ref{eyjcandidate}) cannot all obey (\ref{eqewn}) and (\ref{eqewn2}). (The sole exception is the monogamy of mutual information \cite{mmiref}, i.e. inequality~(\ref{cyclic}) with $k=1$, which can be proven without invoking nesting-violating candidate wedges.) A special class of instances where the tension between (\ref{defcontraction}, \ref{defbc}) and wedge nesting is most easily seen is given by equations~(\ref{sumcoeff}) and (\ref{xeasy}) below, but the phenomenon occurs more generally. 

We spell out the conditions under which candidate wedge $E(Y_j)^{\rm candidate}$ violates (\ref{eqewn}) or (\ref{eqewn2}):
\begin{align}
Y_j \subset X_i & \quad {\rm and} \quad f(x)_j = 1 \quad {\rm and} \quad x_i = 0 \label{violation1} \\
Y_j \cap X_i = \emptyset & \quad {\rm and} \quad f(x)_j = 1 \quad {\rm and} \quad x_i = 1 \label{violation2} \\
Y_j \supset X_i & \quad {\rm and} \quad f(x)_j = 0 \quad {\rm and} \quad x_i = 1 \label{violation3}
\end{align}
These conditions follow from substituting (\ref{eyjcandidate}) for $E(Y_j)$ in the wedge nesting theorems (\ref{eqewn}) and (\ref{eqewn2}). 

Observe that whenever circumstances~(\ref{violation1}-\ref{violation3}) occur and $W(x)$ is non-empty, $E(Y_j)^{\rm candidate}$ is necessarily distinct from the physical entanglement wedge $E(Y_j)$. To wit, when (\ref{violation1}) or (\ref{violation2}) applies, including a non-empty $W(x)$ in $E(Y_j)^{\rm candidate}$ according to (\ref{eyjcandidate}) causes the latter to violate entanglement wedge nesting, which the physical $E(Y_j)$ is guaranteed to respect. Likewise, when (\ref{violation3}) applies, excluding a non-empty $W(x)$ from $E(Y_j)^{\rm candidate}$ according to (\ref{eyjcandidate}) forces the latter to violate entanglement wedge nesting, which the physical $E(Y_j)$ respects. 
\smallskip

\textit{Argument and result:---} Consider a holographic entropy inequality~(\ref{schematic}) and a contraction map $f$, which proves it. 
\begin{enumerate}
\item If $x$ satisfies one of conditions~(\ref{violation1}-\ref{violation3}) and $W(x)$ is non-empty then $E(Y_j)^{\rm candidate} \neq E(Y_j)$.
\item If (\ref{schematic}) is saturated then we have equality in step $(\sharp)$ in (\ref{schematic2}) and, therefore, $E(Y_j)^{\rm candidate} = E(Y_j)$.
\item Thus, if $x$ satisfies one of conditions~(\ref{violation1}-\ref{violation3}) and inequality~(\ref{schematic}) is saturated then $W(x) = \emptyset$. 
\end{enumerate}
Point 3. above is our main result. When introducing regions $W(x)$, we pointed out that the bit string $x$ lists which boundary erasures are protected against in the holographic erasure correcting code. Thus, the non-saturation of inequality~(\ref{schematic}) is a necessary condition for the scheme of erasure correction characterized by $x$. 

In step 2. above, equality in (\ref{schematic2})($\sharp$) directly implies $S_{Y_j}^{\rm candidate} = S_{Y_j}$ for all $j$ because the latter are global minima. One might worry that $E(Y_j) \neq E(Y_j)^{\rm candidate}$ even though $S_{Y_j}^{\rm candidate} = S_{Y_j}$. But this could only occur if the entanglement wedge of $Y_j$ were right at a phase transition between $E(Y_j)^{\rm candidate}$ and $E(Y_j)$, a circumstance in which the $S_{Y_j}$ incur enhanced geometric corrections \cite{trans1, trans2}. In Supplemental Material \cite{sm} we show that our main result---if properly applied---continues to hold even in this fine-tuned scenario.
\smallskip

\textit{Applications.---} Applied to the presently known holographic entropy inequalities, our result gives rise to a plethora of necessary conditions for various schemes of erasure correction. In fact, as discussed in detail in \cite{viols}, contraction proofs of \emph{all} known inequalities except the monogamy of mutual information feature multiple instances of (\ref{violation1}-\ref{violation3}). For a generic example, in Supplemental Material \cite{sm} we randomly choose two inequalities from \cite{1866} and inspect particular contraction maps that prove them. We find that saturating the inequality eliminates 340 (resp. 585) out of 569 (resp. 1145) a priori admissible erasure correction schemes.

Our result has broad applications because it relies on the \emph{existence of one} contraction map $f$ with properties~(\ref{violation1}-\ref{violation3}). In particular, we do not need the nesting violation to occur in all contraction maps, only in one. This makes it algorithmically easy to check if a given inequality constrains a specific erasure correcting scheme $x$. One lists all potential choices of $f(x)$ that satisfy (\ref{violation1}) or (\ref{violation2}) or (\ref{violation3})---call them $f_q(x)$---and looks for a contraction map $f$ with $(x, f_q(x))$ imposed as an additional boundary condition. If such an $f$ exists then the saturation of the inequality guarantees that $W(x)$ is empty. 

There are special instances of our result, which one can write down by hand after glimpsing a given inequality \cite{viols}. To do so, fix a subsystem $A$ and rewrite the inequality such that $A$ appears in every term, trading $X_i$ or $Y_j$ for their complements where necessary. In this convention, $x^A = 1\ldots 1$ and $f^A = 1\ldots 1$ (all ones) is the boundary condition (\ref{defbc}) for $A$. Suppose that---after the rewriting---(\ref{schematic}) contains a region $X_{i^*}$ such that the combined coefficient of all its subsets is negative: 
\begin{equation}
\sum_{i\,|\,X_i \subseteq X_{i^*}} \alpha_i - \sum_{j\,|\,Y_j \subseteq X_{i^*}} \beta_j < 0.
\label{sumcoeff}
\end{equation}
Then for $x$ defined by
\begin{equation}
x_i = \begin{cases} 0 & \textrm{if $X_i \subseteq X_{i^*} $}  \\ 1 & \textrm{otherwise} 
\end{cases}
\label{xeasy}
\end{equation}
imposing (\ref{defcontraction}) with $\tilde{x} = x^A$ implies that $f(x)$ must contain at least one instance of (\ref{violation1}). Thus, whenever (\ref{schematic}) is saturated the $W(x)$ defined in (\ref{xeasy}) is empty. Example~(\ref{5partyex}-\ref{region5}) belongs to this class, with $A = A_3$ and $X_{i^*} = A_2^+$. 

\smallskip
\textit{Applications to the infinite families of inequalities.---} 
The necessary conditions on holographic erasure correction imposed by the infinite families of inequalities proven in \cite{ourineqs} are particularly convenient to state and use. Due to the regular structure of the inequalities and the contraction maps, the necessary conditions pertain to intersections of only a few entanglement wedges \cite{viols}; we provide examples with six or seven of them. This is in contrast to a generic holographic inequality, whose nesting violations typically concern specific combinations of all terms on the `greater-than' side of~(\ref{schematic}). 

The toric inequalities are defined for pure states on systems $A_s$ ($1 \leq s \leq 2k_A+1$) and $B_t$ ($1 \leq t \leq 2k_B+1$) where---like in inequalities~(\ref{cyclic})---the indices are cyclically ordered. Extending notation~(\ref{defap}) to consecutively indexed tuples of $B$-type regions, we have:
\begin{equation}
\sum_{s=1}^{2k_A+1}\, \sum_{t=1}^{2k_B+1} \left( S_{A_s^+ B_t^-} - S_{A_s^- B_t^-}\right) \geq S_{A_1 \ldots A_{2k_A+1}}
\label{eqtoric}
\end{equation}
In addition to these \emph{toric} inequalities, we also have the family of \emph{projective plane} inequalities, which are discussed in Supplemental Material \cite{sm}. The ensuing analysis is phrased with reference to~(\ref{eqtoric}) but it applies to the projective plane inequalities without modifications. 

\begin{figure}[t]
    \centering
    \includegraphics[width=0.48\linewidth]{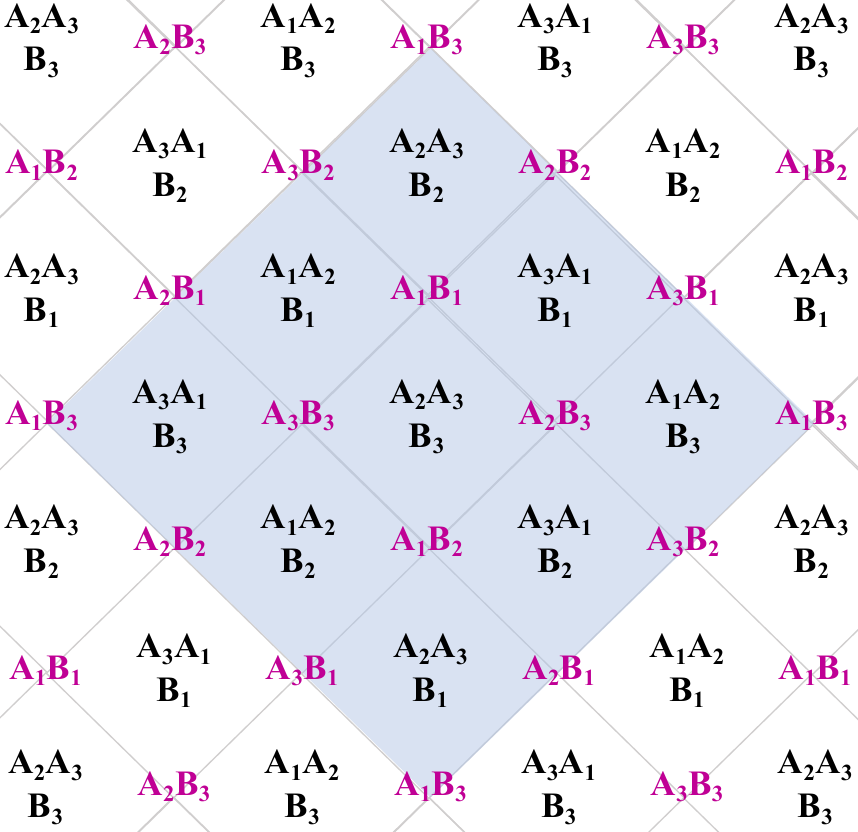}
    \,\,
    \raisebox{0.0\linewidth}{
    \includegraphics[width=0.46\linewidth]{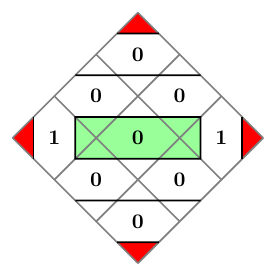}}
    \caption{Left: Terms in the sum in (\ref{eqtoric}) presented as faces and vertices of a graph embedded on a torus. Right: A visual representation of a bit string $x$. Pairs of vertical (resp. horizontal) lines are drawn on faces $X_i$ if the corresponding bit is $x_i = 1$ (resp. $x_i = 0$). We color the interiors of contractible loops; because the green loop is non-minimal, $f(x)$ satisfies one of conditions~(\ref{violation1}-\ref{violation3}).}
    \label{fig:toric}
\end{figure}

Reference~\cite{ourineqs} proved inequalities~(\ref{eqtoric}) by contraction. To conduct the proof, they presented the terms in the double sum in (\ref{eqtoric}) as faces ($X_i \equiv A_s^+ B_t^-$) and vertices ($Y_j \equiv A_s^- B_t^-$) of a graph embedded on a torus; see Figure~\ref{fig:toric}. A vertex and a face are incident if and only if $Y_j \subset X_i$ or $Y_j \cap X_i = \emptyset$ or $Y_j \supset X_j$---that is precisely when the entanglement wedge nesting theorem applies. This has the benefit that instances of (\ref{violation1}-\ref{violation3}) are locally detectable on the graph. 

For a visual representation of bit strings $x \in \{0,1\}^{(2k_A+1)(2k_B+1)}$, on every face $X_i$ draw a pair of vertical (if $x_i = 1$) or horizontal ($x_i = 0$) lines; see Figure~\ref{fig:toric}. With this recipe, setting the $j^{\rm th}$ bit of $f(x)$ to 1 avoids (\ref{violation1}-\ref{violation3}) only if vertex $Y_j$ is surrounded by a minimal loop. In all other circumstances, setting $f(x)_j=1$ incurs a nesting violation and our result applies. 

Yet the contraction proof of \cite{ourineqs} places a single 1 in each collection of vertices $Y_j$, which are surrounded by \emph{any} loop in the representation of $x$. Since only minimal loops avoid nesting violations (\ref{violation1}-\ref{violation3}), we conclude that:
\begin{itemize}
\item Whenever the visual representation of $x$ contains a non-minimal contractible loop, one of conditions~(\ref{violation1}-\ref{violation3}) arises and the saturation of (\ref{eqtoric}) implies $W(x) = \emptyset$.  
\end{itemize}
(Note that non-contractible loops fail to surround vertices.) This result is strong because its applicability is not conditioned on all bits of $x$ but only on those that assemble a non-minimal loop. For example, the loop with the green interior in Figure~\ref{fig:toric} involves only seven bits of $x$; see Supplemental Material \cite{sm} for detailed examples. Moreover, multiple non-minimal contractible loops can be drawn in Figure~\ref{fig:toric}, so the saturation of (\ref{eqtoric}) constrains many distinct erasure correcting schemes $x$. 

\smallskip
\textit{Discussion.---} To our knowledge, these results are the first instance in which holographic entropy inequalities are interpreted in terms of a task---in this case, erasure correction. Prior work used the $k=1$ inequality (\ref{cyclic}), i.e. the monogamy of mutual information \cite{mmiref}, for detecting topological order \cite{toporder} and for amplitude computations \cite{abhy}, but other inequalities have been harder to interpret \cite{arrangement, marginals, hcae}. A loosely related result is that---subject to a genericity assumption---the entanglement wedge of a composite system is connected iff a certain combination of its constituents' entanglement entropies vanishes \cite{1866}. Unlike~\cite{1866}, we focus on holographic entropy inequalities~(\ref{schematic}), exploit properties of contraction maps, and interpret the result in terms of erasure correction.

The Ryu-Takayanagi prescription is subject to quantum corrections from entanglement among bulk (logical) degrees of freedom \cite{qes}. We have ignored them in this paper because our motivation is to design holographic codes, which protect logical information from specific erasures. From this viewpoint, it is never helpful to entangle two logical degrees of freedom before encoding and protecting them against distinct erasure types---which is how bulk entanglement might contaminate our results. In other words, we consider applying holographic erasure correcting codes to product states in the logical Hilbert space and, therefore, all entropies in this paper should be understood as surface areas.

While we have assumed a time reversal-symmetric setup in this work, it is interesting to contemplate how dynamics may affect our result. This offers a potentially new perspective on the still unsettled question of whether holographic entropy inequalities for time-dependent states are weaker than those for static states \cite{withxi, mattveronika, mattveronikanew}. Our result also applies to entanglement wedges of bulk regions \cite{bp1, bp2} because the latter obey all holographic entropy inequalities proven by contraction \cite{hecforbulk}. 

\begin{acknowledgements}
\noindent
We thank Bowen Chen, Ricardo Esp{\'i}ndola, Keiichiro Furuya, Yichen Feng, Minjun Xie and Dachen Zhang for discussions. BC thanks the organizers of the \emph{Observables in Quantum Gravity: From Theory to Experiment} program, which was held at the Aspen Center for Physics and supported by National Science Foundation grant PHY-2210452, where some of this work was carried out. Some figures were previously used in References~\cite{ourineqs, viols} of which we are co-authors. YW is supported by the Shuimu Tsinghua Scholar Program of Tsinghua University. This research was supported by an NSFC grant number 12042505 and a BJNSF grant under the Gao Cengci Rencai Zizhu program.

\end{acknowledgements}

\newpage
$\phantom{.}$
\newpage
\section*{Supplemental Material}

\subsection{Our claim at a phase transition}
\label{app:a}
\noindent
In step 2. of the \emph{Argument} in the main text, we claimed that
\begin{equation}
\sum_{j=1}^r \beta_j\, S_{Y_j}^{\rm candidate} 
= \sum_{j=1}^r \beta_j\, S_{Y_j}
\label{appsaturate}
\end{equation}
implies $E(Y_j)^{\rm candidate} = E(Y_j)$ for all $j$. The direct consequence of (\ref{appsaturate}) is that $S_{Y_j}^{\rm candidate} = S_{Y_j}$ for all $j$ because the $S_{Y_j}$ are global minima. To complete the argument, we inspect the possibility that $S_{Y_j}^{\rm candidate} = S_{Y_j}$ yet $E(Y_j)^{\rm candidate} \neq E(Y_j)$ for some $j$. 

If $S_{Y_j}^{\rm candidate} = S_{Y_j}$ then, in fact, the surfaces that bound $E(Y_j)^{\rm candidate}$ and $E(Y_j)$ are both global minima so we are at a phase transition for the entanglement wedge of $Y_j$. As we explain below, this implies that the entanglement wedge of some $X_i$ is also at a phase transition, so regions $W(x)$ to which our claim pertains are not well defined. Moreover, such a setup is problematic from the viewpoint of designing erasure correction codes: it means that the erasure protection of some bulk degrees of freedom is marginally unstable in the sense that arbitrarily small perturbations of the state can invalidate it. Even if we ignore this problem and---as an academic exercise---attempt to construct an erasure correcting code right at a holographic phase transition, we find that our core result still applies so long as the regions $W(x)$ are consistently defined.

We are interested in the circumstance that $E(Y_j)^{\rm candidate}$ violates entanglement wedge nesting. For definiteness, we consider a nesting violation of type~(\ref{violation1}) from the main text, that is where $E(Y_j)^{\rm candidate}$ includes a non-empty $W(x)$, which is outside the entanglement wedge of some $X_i \supset Y_j$. (The reasoning that applies to the other nesting violations---scenarios~(\ref{violation2}) and (\ref{violation3}) in the main text---is nearly identical.) Our task is to inspect the possibility that there is some `correct' entanglement wedge $E(Y_j)$, which is distinct from $E(Y_j)^{\rm candidate}$ and which respects nesting: $E(Y_j) \subset E(X_i)$.

\begin{figure}
    \centering
    \includegraphics[width=0.98\linewidth]{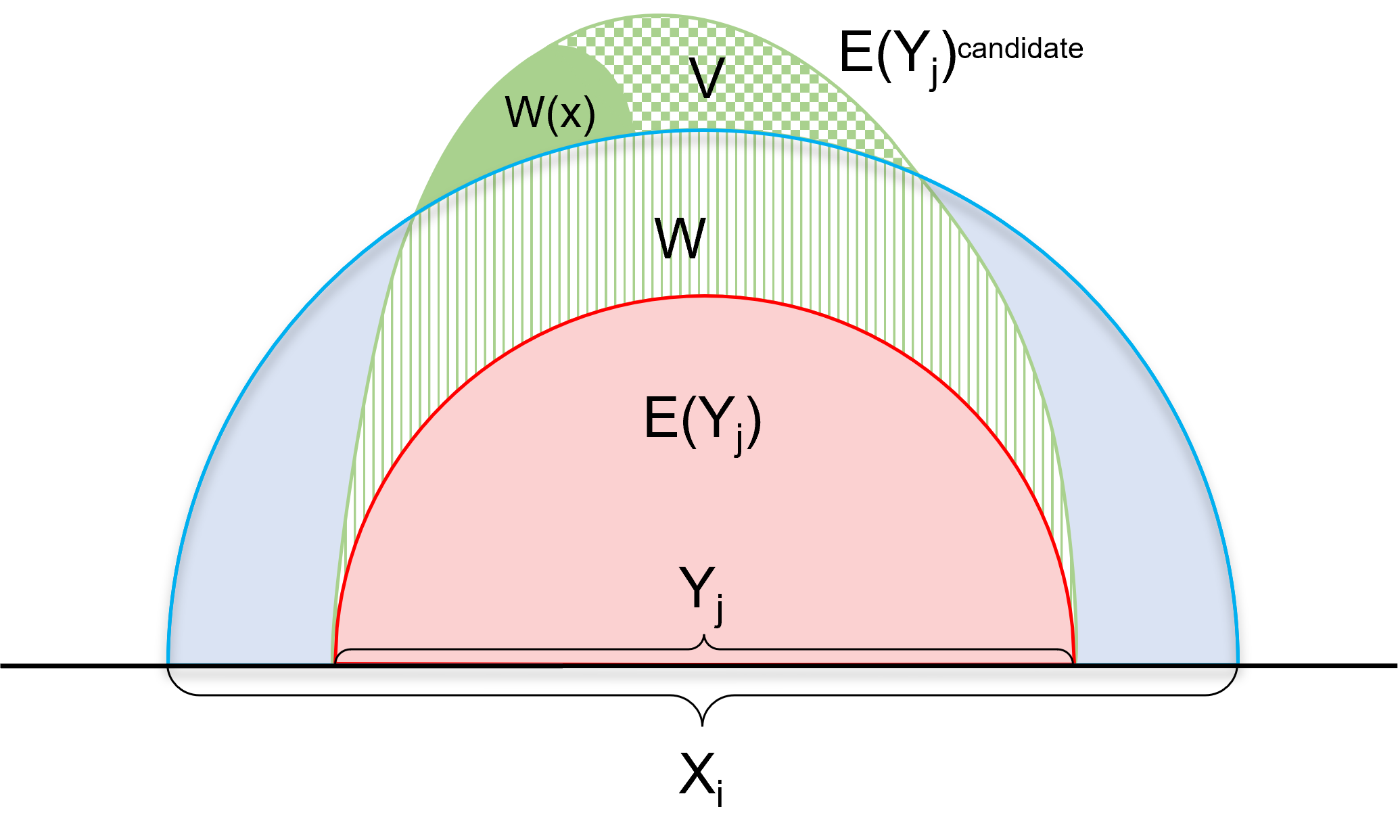}
    \includegraphics[width=0.7\linewidth]{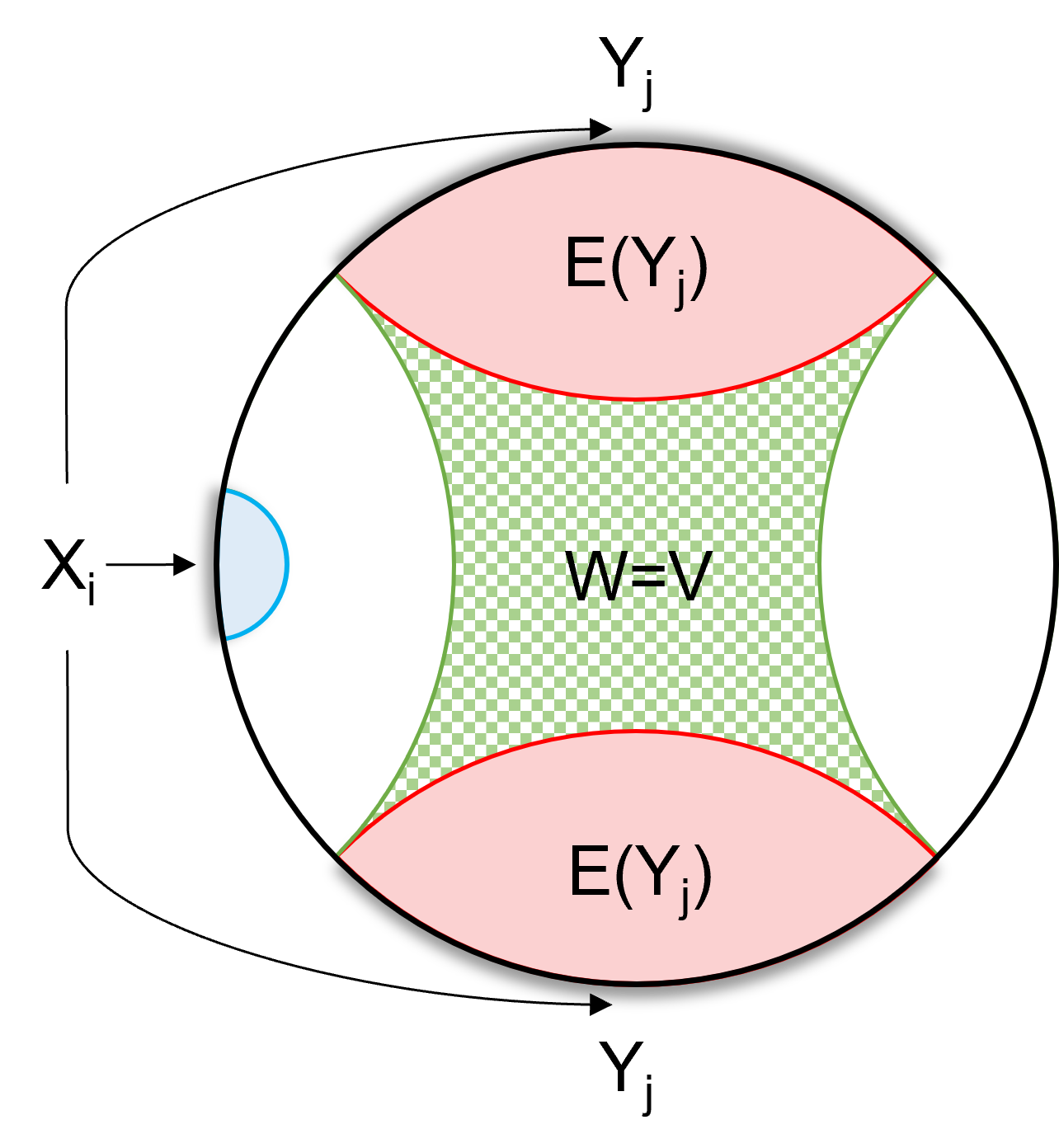}
    \caption{Upper panel: Definitions in the argument in Section\ref{app:a}. $W$ comprises all green regions while $V$ comprises the darker-green regions, including $W(x)$. Lower panel: A realistic scenario where the argument may apply, with $E(Y_j)$ at a phase transition.}
    \label{fig:pt}
\end{figure}

The argument is illustrated in Figure~\ref{fig:pt}. Let $W \equiv E(Y_j)^{\rm candidate} \setminus E(Y_j)$; we have $W(x) \subset W$. Note that $W$ does not touch the boundary or else including it in $E(Y_j)^{\rm candidate}$ would affect the homology condition for $Y_j$. Define $V \equiv W \setminus E(X_i)$; $V$ is non-empty because it also contains $W(x)$. By construction, $\partial V$ has two parts: one in $\partial E(X_i)$ and one in $\partial E(Y_j)^{\rm candidate}$. Observe that
\begin{equation*}
{\rm Area}\big(\partial E(X_i) \cap \partial V\big) = {\rm Area}\big(\partial E(Y_j)^{\rm candidate} \cap \partial V\big)
\end{equation*}
or else you could trade one for the other in the computation of $S_{X_i}$ or $S_{Y_j}^{\rm candidate}$ and decrease the quantity. This means that $V$ may or may not be included in the entanglement wedge of $X_i$ without affecting the entropy, so the said wedge is indeed at a phase transition. 

If the entanglement wedge of $X_i$ is at a phase transition then bulk regions $W(x)$ are not well-defined until we specify which of the equal-area Ryu-Takayanagi surfaces bounds what we call $E(X_i)$. In the present case, $E(Y_j)^{\rm candidate}$ appears to violate entanglement wedge nesting because it juts out of $E(X_i)$ even though $Y_j \subset X_i$. But this is only because the phase transition allows us to choose what we call the entanglement wedge of $Y_j$ and $X_i$, and we made the larger choice (with $W$ in it) for $Y_j$ but the smaller choice (without $V$) for $X_i$. To be consistent, if we work with the larger choice of $E(Y_j)^{\rm candidate}$ then we should also include the contentious region $V$ in the entanglement wedge of $X_i$.

Because $W(x) \subset V$, including $V$ in the entanglement wedge of $X_i$ flips the $i^{\rm th}$ bit of $x$ from 0 to 1 and condition~(\ref{violation1}) from the main text no longer applies. Thus, our claim remains valid even when the holographic erasure correcting code sits right at a phase transition. This is so long as the bulk regions $W(x)$ are defined appropriately---that is, by taking intersections of entanglement wedges in consistently chosen phases. 

\subsection{Example application}
\noindent
For a sense of how our result can be applied in practice, we inspect two randomly chosen holographic entropy inequalities from \cite{1866}---number 1311 and 1722. For each of them, we generate one contraction map that proves it using \cite{contractor} and identify instances of (\ref{violation1}-\ref{violation3}). The analyses are summarized in Microsoft Excel files, which are included with the arXiv submission of this paper.

The short summary is that the saturation of the relevant inequalities excludes roughly half of potentially admissible schemes of erasure correction.
\smallskip 

\textit{Inequality~1311.---} The inequality reads:
\begin{align}
2 S_{ABC} + S_{ABD}\, +\, &S_{ABE} + S_{ABF} + S_{ACD} \nonumber \\
+\, S_{ADE} + 2 S_{BCD} \,+\, &2S_{BDE} + S_{BEF} + S_{CDE} \nonumber \\
& \geq \label{1311} \\
S_A + S_{AB} + S_{AC} + S_{BC}\, +\, &S_{BD} + S_{BE} + S_{BF} + S_{CD} \nonumber \\
+\, 2S_{DE} + 2S_{ABCD} +S&_{ABDE} + S_{ABEF} + S_{BCDE} \nonumber
\nonumber
\end{align}
There are $l = 10$ distinct terms on the `greater-than' (LHS) side so, na\"{\i}vely, there are $2^{10} = 1024$ bit strings $x \in \{0,1\}^l$. In fact, there are only
\begin{equation}
\sfrac{3}{4} \times \sfrac{3}{4} \times 1024 = 576
\end{equation}
physically sensible bit strings because combinations
\begin{align}
x_{ABF} = x_{CDE} = 1 \label{nonsense1a} \\
x_{ACD} = x_{BEF} = 1 \label{nonsense1b}
\end{align}
violate entanglement wedge nesting (disjoint systems would have overlapping entanglement wedges). The contraction map need not be defined on the $1024 - 576 = 448$ bit strings indicated in (\ref{nonsense1a}-\ref{nonsense1b}) as the underlying regions $W(x)$ are guaranteed ahead of time to be empty \cite{physstrings}. In what follows we only consider the 576 physically admissible bit strings $x$.

Of the 576 physically admissible erasure correction schemes, seven are boundary conditions~(\ref{defbc}) so their existence is not under question. Among the remaining 569 bit strings $x$, 340 exhibit nesting violations~(\ref{violation1}-\ref{violation3}) under the contraction map $f$ we generated. When~(\ref{1311}) is saturated, the corresponding regions $W(x)$ are empty.

We give one specific instance of a bit string $x$ to which our result applies:
\begin{align}
x_{BDE} = x_{CDE} &= 1 \nonumber \\
x_{ABC} = x_{ABD} = x_{ABE} = x_{ABF} & \label{onex1311} \\
= x_{ACD} = x_{ADE} = x_{BCD} = x_{BEF} & = 0 \nonumber
\end{align}
That is, if (\ref{1311}) is saturated then a holographic erasure correcting code cannot protect logical information against the erasure of $ABC$, $ABD$, $ABE$, $ABF$, $ACD$, $ADE$, $BCD$ and $BEF$ while keeping it accessible from $BDE$ and $CDE$. 

This example is somewhat intriguing. Described in broad strokes, it looks like the erasure of $A$ is well protected against while $DE$ is very helpful but not sufficient for recovery. Overall, (\ref{onex1311}) looks like a reasonable set of requirements. We find it remarkable that the non-existence of such a holographic erasure correcting code can be intuited from the saturation of~(\ref{1311}). 

There is one reason why one might expect our result to be particularly strong when applied to inequality~(\ref{1311}): a single-region entropy $S_A$ appears on the `less-than' (RHS) side. This might enhance the strength of our result because a single-region term trivially satisfies the set-inclusion conditions in (\ref{violation1}-\ref{violation2}) with all LHS terms $X_i$. In (\ref{1311}), we obviously have either $A \subset X_i$ or $A \cap \overline{X_i} = \emptyset$ for all $i$. This suggests that wedge nesting violations~(\ref{violation1}-\ref{violation2}) might be particularly abundant for inequalities, which feature single region terms. 

For this reason, we randomly picked one other inequality from \cite{1866}, imposing an extra demand that it contain no single-region terms.  The goal of this exercise is to filter out any potential enhancement of our result, which is introduced by the appearance of a single region term.
\smallskip

\textit{Inequality~1722.---} The inequality reads:
\begin{align}
S_{ABC} + S_{ABD}\, +\,  &2S_{ABE} + S_{ACD} + S_{ADF} + S_{BDE} \nonumber \\
+\, S_{ACEF} + S_{ADEF} \,+\, & S_{BCDF} + S_{BCEF} + S_{BDEF} \nonumber \\
\geq & \label{1722} \\
S_{AB} + S_{AC} + S_{AD}\, +\, & S_{AE} + S_{BC} + S_{BD} + S_{BE} \nonumber \\
+\, S_{DF} \,+\, & S_{EF} \nonumber \\
+\, S_{ABDE} +S_{ACDF} \,+\, & S_{ABCEF} +  S_{ABDEF} + S_{BCDEF} \nonumber
\nonumber
\end{align}
There are $l = 11$ distinct terms on the `greater-than' (LHS) side so, na\"{\i}vely, there are $2^{11} = 2048$ bit strings $x \in \{0,1\}^l$. In fact, there are only
\begin{equation}
\sfrac{3}{4} \times \sfrac{3}{4} \times 2048 = 1152
\end{equation}
physically sensible bit strings because combinations
\begin{align}
x_{ADF} = 1 \qquad & {\rm yet} \qquad x_{ADEF} = 0\label{nonsense2a} \\
x_{BDE} = 1 \qquad & {\rm yet} \qquad x_{BDEF} = 0 \label{nonsense2b}
\end{align}
violate entanglement wedge nesting. The $2048 - 1152 = 896$ bit strings indicated in (\ref{nonsense2a}-\ref{nonsense2b}) are automatically empty and the contraction map need not be defined on them \cite{physstrings}. We only consider the 1152 physically admissible bit strings $x$.

Of the 1152 physically admissible erasure correction schemes, seven are boundary conditions~(\ref{defbc}) so their existence is not under question. Among the remaining 1145 bit strings $x$, 569 exhibit nesting violations~(\ref{violation1}-\ref{violation3}) under the contraction map $f$ we generated. When~(\ref{1722}) is saturated, the corresponding regions $W(x)$ are empty.

\subsection{Application to projective plane inequalities}
\textit{Preliminaries.---} 
Projective plane inequalities are defined for pure states on systems $A_s$ and $B_t$, with $1 \leq s, t \leq m$. The indices $s$ and $t$ are again cyclically understood, i.e. $s \equiv s + m$. Note that the number of $A$- and $B$-type systems is equal and can be either odd or even.

\begin{figure*}
    \centering
    \raisebox{0.005\linewidth}{
    \includegraphics[width=0.48\linewidth]{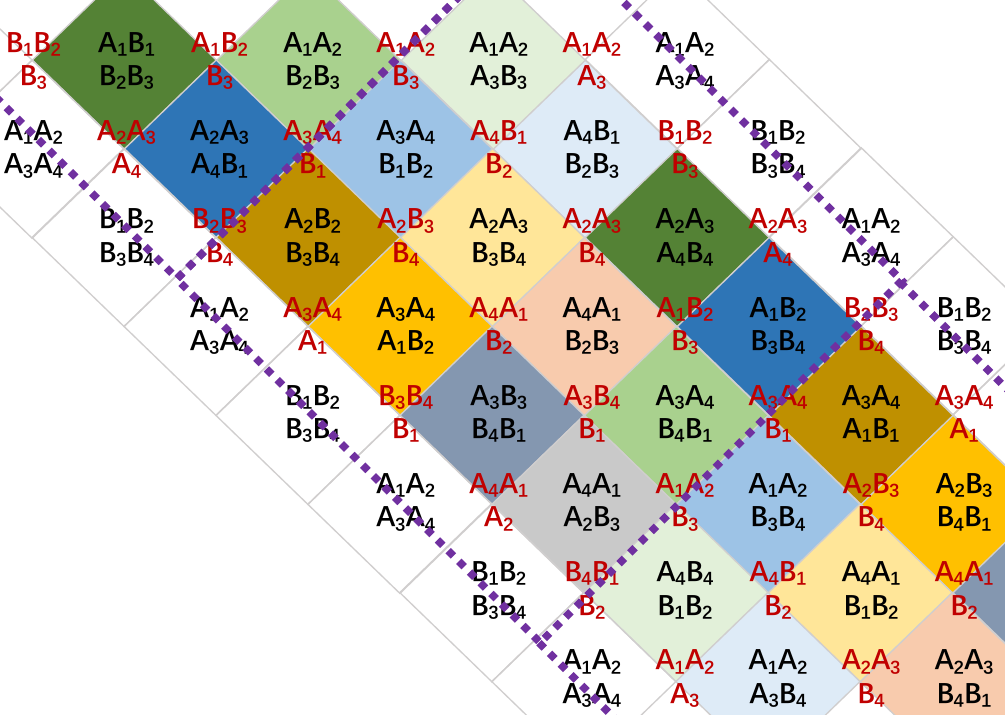}}\,\,
    \raisebox{0.0\linewidth}{
    \includegraphics[width=0.48\linewidth]{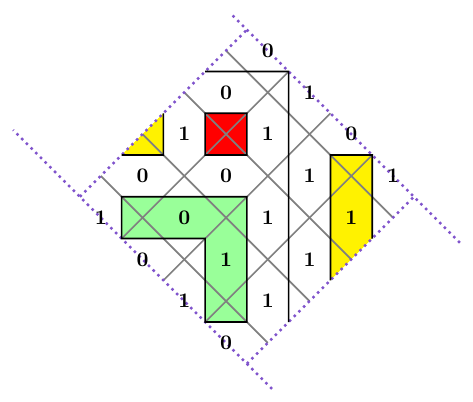}}
    \caption{Left: The graph, which represents the projective plane inequalities~(\ref{rp2ineqs}), here shown for $m=4$. Terms on the `greater-than' side are faces and terms on the `less-than' side are vertices; we color-code the faces to highlight multiple appearances of the same face. The special term $S_{A_1 A_2 \ldots A_m} = S_{B_1 B_2 \ldots B_m}$ appears as the white squares on the extreme diagonals; those squares should be understood as one face. The graph has the topology of the M{\"o}bius strip glued to a disk (which represents $S_{A_1 A_2 \ldots A_m}$), which is why inequalities~(\ref{rp2ineqs}) are called `projective plane inequalities.' 
   \\$\phantom{~~~~~~~~~~}$
    Right: A visual representation of an example bit string $x \in \{0,1\}^l$. A pair of vertical (respectively, horizontal) lines is drawn on faces whose corresponding bits equal 1 (respectively 0) in $x$. The visual representation is made up of loops. We color the interiors of the topologically trivial (contractible) loops; in this example the red loop is minimal whereas the green and yellow loops are non-minimal. In the contraction map given in \cite{ourineqs}, every non-minimal contractible loop gives rise to a nesting violation (\ref{violation1}-\ref{violation3}). Therefore, our result applies to all sequences of bits, which make up non-minimal contractible loops like the yellow and green loop shown here.
    }
    \label{fig:rp}
\end{figure*}

We introduce the notation
\begin{equation}
A_s^{(p)} = A_s A_{s+1} \ldots A_{s+p-1} 
\end{equation}
and similarly for $B_t^{(p)}$. Note that the combinations defined in equations~(\ref{defap}-\ref{defam}) in the main text are special cases of this notation: $A_s^+ = A_s^{(k+1)}$ and $A_s^- = A_s^{(k)}$. 

The projective plane inequalities take the form \cite{ourineqs}:
\begin{align} 
\!(m-1) & S_{A_1A_2\ldots A_m}\! + \frac{1}{2}\! \sum_{t=1}^{m-1} \!\sum_{s=1}^{m}\!
\left(S_{A_s^{(t)} B_{s+t-1}^{(m-t)}} \!+\! S_{A_s^{(t)} B_{s+t}^{(m-t)}} \!\right) \nonumber \\ 
& \geq 
\sum_{s,t=1}^{m} S_{A_s^{(t-1)} B_{s+t-1}^{(m-t)}}
\label{rp2ineqs}
\end{align}
The sum on the `greater-than' side of the inequality ostensibly carries coefficient $\sfrac{1}{2}$ but, in fact, every term appears twice in the sum---once as the entropy of a region $X$ and once as that of $\overline{X}$. Therefore, all \emph{distinct} terms other than $S_{A_1 A_2 \ldots A_m}$ actually have unit coefficient. 

To streamline the analysis, we again present the terms of the inequality as faces and vertices of a graph. Distinct terms on the `greater-than' side are faces while terms on the `less-than' side are vertices. A vertex and a face are incident if and only if they satisfy the premise of the entanglement wedge nesting theorem, that is if they are in one of the relations listed in conditions~(\ref{violation1}-\ref{violation3}) in the main text. This guarantees that any and all violations of entanglement wedge nesting are locally detectable on the graph. The resulting graph is shown in the left panel of Figure~\ref{fig:rp}. It has the topology of the projective plane, which explains the nomenclature for inequalities~(\ref{rp2ineqs}). Notice the similarity with the left panel in Figure~\ref{fig:toric} in the main text.

Proceeding as we did with the toric inequalities in the main text, we represent bit strings $x \in \{0, 1\}^l$ on the graph by drawing a pair of vertical (respectively, horizontal) lines on a face if its corresponding bit in $x$ is $x_i = 1$ (respectively, $x_i = 0$); see the right panel of Figure~\ref{fig:rp}. This produces a set of loops on the graph, all but one of which are contractible \cite{ourineqs}.

The present setup shares a key structural property with the discussion of the toric inequalities from the main text:
\begin{itemize}
\item Given a bit string $x$, a contraction map can assign 1 to a vertex without a nesting violation (\ref{violation1}-\ref{violation3}) only if it is surrounded by a minimal loop.
\end{itemize}
This follows directly from applying conditions (\ref{violation1}-\ref{violation3}) to the definitions above.
\smallskip

\textit{Contraction map.---} 
Reference~\cite{ourineqs} proved inequalities~(\ref{rp2ineqs}) by producing an explicit contraction map. Their map has the following property: 
\begin{itemize}
\item Whenever a collection of vertices is surrounded by a contractible loop, the map in \cite{ourineqs} assigns 1 to at least of them. 
\end{itemize}

Combining the two highlighted points, we reach the conclusion:
\begin{itemize}
\item Whenever the visual representation of a bit string $x$ contains a non-minimal contractible loop, one of conditions~(\ref{violation1}-\ref{violation3}) occurs. Therefore, for such $x$, if (\ref{rp2ineqs}) is saturated then $W(x) =\emptyset$. 
\end{itemize}
This is identical to how our result applies to the toric inequalities~(\ref{eqtoric}) from the main text. We emphasize that this class of applications is stronger than generic instances of our result because it applies when only a \emph{subset} of the bits of $x$ satisfy a condition. The condition is that a non-minimal contractible loop appear in the analogue of the right panel of Figure~\ref{fig:rp}.
\smallskip

\textit{Examples.---}
We consider the bit string $x$ from Figure~\ref{fig:rp} for illustration.

The green loop expresses a particular combination of seven bits of $x$. By comparing the two panels in the figure, we see that they are:
\begin{align}
x_{A_1 A_2 A_3 A_4} & = 1 \nonumber \\
x_{A_2 B_2 B_3 B_4} & = 0 \nonumber \\
x_{A_3 A_4 A_1 B_2} & = 0 \nonumber \\
x_{A_2 A_3 B_3 B_4} & = 0 \\
x_{A_4 A_1 B_2 B_3} & = 1 \nonumber \\
x_{A_3 B_3 B_4 B_1} & = 1 \nonumber \\
x_{A_4 A_1 A_2 B_3} & = 1 \nonumber 
\end{align}
The green loop tells us that if inequality~(\ref{rp2ineqs}) at $m=4$ is saturated then the holographic erasure correcting code cannot simultaneously protect any logical information from the erasure of $B_1 B_2 B_3 B_4$ and $A_2 B_2 B_3 B_4$ and $A_3 A_4 A_1 B_2$ and $A_2 A_3 B_3 B_4$ and $A_2 A_3 B_4 B_1$ and $A_4 A_1 A_2 B_2$ and $A_3 B_4 B_1 B_2$. (Note that the code protects against the erasure of systems for which $x_i = 0$. For bits where $x_i = 1$, the code protects against the erasure of their complements.)

Similarly, the yellow loop in the right panel of Figure~\ref{fig:rp} expresses a particular combination of six bits of $x$. By comparing the two panels, we read off:
\begin{align}
x_{A_1 A_2 A_3 A_4} & = 1 \nonumber \\
x_{A_2 A_3 A_4 B_4} & = 1 \nonumber \\
x_{A_1 B_2 B_3 B_4} & = 1 \\
x_{A_3 A_4 B_4 B_1} & = 1 \nonumber \\
x_{A_3 A_4 B_1 B_2} & = 1 \nonumber \\
x_{A_2 B_2 B_3 B_4} & = 0 \nonumber
\end{align}
The yellow loop tells us that if inequality~(\ref{rp2ineqs}) at $m=4$ is saturated then the holographic erasure correcting code cannot simultaneously protect any logical information from the erasure of $B_1 B_2 B_3 B_4$ and $A_1 B_1 B_2 B_3$ and $A_2 A_3 A_4 B_1$ and $A_1 A_2 B_2 B_3$ and $A_1 A_2 B_3 B_4$ and $A_2 B_2 B_3 B_4$.
\end{document}